\documentclass[aps,amsfonts,amsmath,prd,preprint,nofootinbib]{revtex4}
\newcommand{\beq}{\begin{equation}}
\newcommand{\eeq}{\end{equation}}

\usepackage{epsfig}
\usepackage{bbm,cancel,ulem}
\usepackage{hyperref}
\usepackage{xcolor}
\usepackage{wasysym}
\usepackage{graphicx}
\usepackage{color}
\usepackage{pgfplots}
\pgfplotsset{compat=1.15}
\usepackage{mathrsfs}
\usetikzlibrary{arrows}
\begin{document}
	
	\title{An Ultra-compact Object from Semi-classical Gravity}
	
	\author{I. Prasetyo}
	\email{ilham.prasetyo@sci.ui.ac.id}
	\author{H. S. Ramadhan}
	\email{hramad@sci.ui.ac.id}
	\author{A. Sulaksono}
	\email{anto.sulaksono@sci.ui.ac.id}
	\affiliation{Departemen Fisika, FMIPA, Universitas Indonesia, Depok 16424, Indonesia. }
	
	\def\changenote#1{\footnote{\bf #1}}
	
\begin{abstract}
In a recent report, Carballo-Rubio~\cite{Carballo-Rubio:2017tlh} utilizes the semi-classical theory of gravity to obtain a generalized Tolman-Oppenheimer-Volkoff (TOV) equation. 
This model has a new coupling constant $l_p$ implying two different modified TOV equation forms characterized by the sign of $p'$. 
The negative branch reduces to the ordinary GR-TOV in the limit of $l_p\to0$, while the positive one does not. 
In the positive branch, Carballo-Rubio was able to find the exact solutions using the constant-$\lambda$ trick. In this work, we investigate whether this theory's negative branch can also provide a different feature of the ultra-compact object compared to those obtained from the GR-TOV equation.  We use a numerical method to calculate the properties of an ultra-compact object. We tested our code by calculating numerically  the pressure profile of the object in a positive branch with the star's radius $R$ in a macroscopic unit by integrating from surface to center. The result reproduces the analytical result of  Carballo-Rubio~\cite{Carballo-Rubio:2017tlh}. While in a negative branch, we reproduce the pressure profile results of Ho and Matsuo~\cite{Ho:2017vgi} using constant energy density $\rho=\rho_0$ not only by integrating from surface to center but also by integrating from center to surface, respectively. In this work, we study ultra-compact objects with an isotropically ideal fluid matter where we use a simple but physically motivated equation of state $\rho=p/w+\rho_0$ with $w$=1 and $w$=1/3. In general, we obtain that the range of $l_p$ is very restricted and must not be equal to $r_c$. Here $r_c$ is the starting point of integration located at the center of the star. While $l_p$ should be set to be much larger than Planck length $L_\text{Pl}$. Consequently, the mass-radius curves for the various value of $l_p$ for both  $w$ cases are still indistinguishable from the standard GR-TOV results.   Hence from the negative branch of $p'(r)$, the additional free parameter $l_p$ does not provide a significant effect compared to the standard GR-TOV equation results, even though $l_p$ is not in the limit of $l_p\to0$ anymore. Therefore, similar to the conclusion in Ref.~\cite{Urbano:2018nrs} with GR theory that the ultra-compact objects from negative branch of semi-classical gravity with a linear equation of state are unable to generate demanding gravitational echoes.
\end{abstract}

	\maketitle
	\thispagestyle{empty}
	\setcounter{page}{1}

	\section{Introduction}
	\label{intro}
 
 The recent excitement over gravitational wave astronomy by LIGO and Virgo~\cite{GBM:2017lvd} gave rise to many discussions over what kind of massive bodies produce the waves. From the astronomical point of view, the curiosity of synthesizing heavy metals such as silver and platinum from a collision of two neutron stars also triggered excitement. From the theoretical point of view, many attempts have been made to justify the existence of horizonless ultra-compact objects other than known compact objects such as neutron stars or white dwarfs when interacting with other massive bodies by probing its behavior through its waves. There are many proposals given, including gravastars~\cite{Mazur:2001fv,Mazur:2004fk} where inside the surface of the star lies another de-Sitter space connected to the usual Schwarzschild space-time by a thin shell of ultrarelativistic matter. Interestingly recently, Carballo-Rubio~\cite{Carballo-Rubio:2017tlh} has claimed to predict exotic stars that have many similar features with gravastar using semiclassical gravity theory. 

The semiclassical theory of gravity is probably one of the earliest attempts by physicists to reconcile quantum field theory and gravity. It began with the remarkable result on black hole entropy and temperature and the Unruh effect (see the famous classic monograph~\cite{Birrell:1982ix} and for more recent one see~\cite{Parker:2009uva}). In this theory, the concept of particle creation in general space-time is different from the usual one in Minkowski space-time. This ambiguity is mainly caused by the Bogolyubov transformation, which is similar to coordinate transformation. Although the problem seems unsolvable, it turns out that the energy-momentum tensor is indifferent to what ``coordinate system'' we chose. Even though there are infinities inside them, physicists had identified what kind of infinities leads to the physical quantum state, resulting in the so-called Hadamard state, i.e., a state whose two-point function has infinities whose form is the same as the one calculated in Minkowskian space-time. Another state exists, i.e., Boulware state, which had been extensively studied again since it describes a Schwarzschild black hole vacuum. Since the study of $n$-point functions are related to regularization of infinities from expectation value of energy-momentum tensor, $\langle \hat{T}_{\mu\nu}\rangle$, hence the main interest in the semiclassical theory of gravity is mainly around renormalization of the energy-momentum tensor which is related to the Einstein tensor by
	\[ G_{\mu\nu}=8\pi G \langle \hat{T}_{\mu\nu} \rangle.\]
The resulting renormalization should satisfy Wald's renormalization axioms (see page 89 of~\cite{Wald:1995yp}).  It has been known that the methods of renormalization only can apply for some specific cases~\cite{DeWitt:1975ys}, and in the case of conformal field theory, these methods give rise to the so-called trace anomaly~\cite{Birrell:1982ix} which is a still unsolvable problem in semi-classical gravity theory. The value of the cosmological constant is restricted in a nontrivial way~\cite{Juarez-Aubry:2019jbw}; hence one cannot treat this constant as a free parameter in this theory.

	\textcolor{black}{Here we provide a short discussion related to Boulware and Unruh vacuums.  The definition of Boulware vacuum is a quantum state of the exterior of any massive body but singular at the horizon. In contrast, the Unruh vacuum is the quantum state of gravitational collapse, and it features Hawking radiation at large distances ~\cite{Visser1997,Barbado2011}.  In general, these two vacuums are inequivalent even though they could be interlinked classically by coordinate transformation.  Please see Ref.~\cite{Shapiro2008} for a detailed discussion about how to relate both quantum vacuums. The authors of Ref.~\cite{Barcelo2008} investigated the dynamical evolution of a collapsed star within semiclassical gravity. They found that in some cases, the trapping horizon is prevented from forming in a semiclassical approach, and the new collapsed objects could exist. These objects make the confrontation with information paradox and run-away endpoint problem unnecessary. They also found that both quantum vacuums can describe the same exterior compact object locally as long as the horizon is unformed. If the horizon is formed, then the Boulware vacuum could no longer be used, and one should use Unruh vacuum to describe the object. Please see detail technical discussions of this matter in Ref.~\cite{Barcelo2008}. A recent study by Ho and Matsuo~\cite{Ho:2017vgi}  has shown that if the compact objects have no singularity nor horizons implies that arbitrary heavy objects can have a physical state in the Boulware vacuum as long as it has stationary state. Furthermore, Ho and Matsuo in other paper~\cite{HoMa2018} have shown that it is physically sensible to consider the Boulware vacuum for any macroscopic radius of a compact object from nonperturbative analysis of the semiclassical Einstein equation. These results ~\cite{Ho:2017vgi,HoMa2018} are contrary to the common belief that Boulware vacuum becomes unphysical if the radius is smaller than the Schwarzchild radius since the energy-momentum tensor is divergent at the  Schwarzchild radius. In this work, we study the compact object without horizon, and the radius is larger than the Schwarzchild radius. Therefore, using a Boulware vacuum is relatively safe.}
    
    Meanwhile, on gravitational wave astronomy, there are some indications that all black holes cannot be observed except by their intense gravity effects. Therefore, there exists the possibility of the existence of the so-called black hole mimickers. These are compact horizonless astronomical objects, usually assumed to have spherical symmetry, whose mass is enormous, yet their size is relatively small. As an illustration, a neutron star's mass $M$ had been observed to be around $1.4-2$ solar mass, and its radius $R$ is roughly within the range of 10 to 20 km. These compact objects are mainly featured by their compactness $C=GM/R$, a dimensionless quantity due to Newton's constant $G$.  Those objects can be categorized as follows~\cite{Cardoso:2019rvt}:
	\begin{enumerate}
		\item compact objects, if $2C>1/3$,
		\item ultra-compact objects, if $2C>2/3$,
		\item objects violating Buchdhal limit, if $2C>8/9$,
		\item clean-photon sphere objects, if $2C>1/(1+0.019)$, and
		\item near-horizon quantum structures, if $2C>1/(1+10^{-40})$.
	\end{enumerate} 
A black hole's compactness is $2C=1$, so these exotic compact objects should have compactness less than the black hole's ($2C<1$). Suppose that any two massive objects had collided and formed another massive object in the so-called ring-down phase, this final object, in general, can be anything, including a black hole. Due to the gravitational field's intensity, these final objects will most likely be a black hole, but if it is not, it will be at least an ultra-compact object. By many analyses on a geodesic of light rays propagating near the object, perturbating its exterior gravitational field, people knew that the effective potential $V_{eff}$ from its gravity would affect its ring-down gravitational wave pulse~\cite{Urbano:2018nrs}. After two massive objects are combined into either a black hole or an ultra-compact object, its gravitational wave pulse will stop or produce echoes as the final object forms. This echo is the leaking of those gravitational waves that are trapped by the effective potential barrier located at radius $r=3GM$ ($V_{eff}'(3GM)=0$ and $V_{eff}''(3GM)<0$), which is the location of the so-called photon sphere. To have the photon sphere, then the object's radius should satisfy $R<3GM$, which implies compactness $2C>2/3$. It is worthy to note that the photon sphere's existence is due to the geodesic of light rays propagating near the massive object, but its existence does not imply gravitational wave echo, which came from the so-called TT-tensor part of metric perturbation. The potential wall location at $r=3GM$ is in coincidence with the photon sphere and is a byproduct of the spherical symmetry assumption, which is generally not true for rotating objects that obey axial symmetry. In the rest of this paper, we shall only assume spherical symmetry. The gravitational wave echo frequency $f_{echo}$ is in one-to-one correspondence on the object's compactness $C$ by~\cite{Cardoso:2019rvt} 	

	\begin{equation}
	\tau_{echo}^{(approx)}\sim 4M |\log \epsilon |,\label{eq:tauapprox}
	\end{equation}
	with $\epsilon={1/2C}-1$ and $f_{echo}=\pi/\tau_{echo}$. This echo time delay can also be estimated by a calculation of its metric component $g_{tt}$ and $g_{rr}$ both inside and outside the star~\cite{Mannarelli:2018pjb} 
	\begin{equation}
	\tau_{echo}^{(num)}=\int_{0}^{3GM} \sqrt{-{g_{rr}\over g_{tt}}}~dr.\label{eq:tauapprox0}
	\end{equation}

The integration means that the trapped waves propagated from $r=3GM$ to $r=0$ and back to $r=3GM$ by assuming that the gravitational wave penetrates the ultra-compact object. (In fact, Eq. \eqref{eq:tauapprox} came from Eq. \eqref{eq:tauapprox0} but integrated from $r=R$ to $r=3GM$.) It had been calculated that the binary neutron stars merger GW170817 produced an approximation to a ``tentative'' echo frequency $f_{echo}\simeq72$ Hz from a 2.6-2.7 solar mass "black hole" remnant with dimensionless spin $0.84-0.87$~\cite{Abedi:2018npz}. This observation had been predicted to be compatible with a toy model of an incompressible star with mass $2-3$ solar mass~\cite{Pani:2018flj}. There are varieties of the proposed compact objects (see the compilation of the properties of those objects and the corresponding discussions in Ref.~\cite{Cardoso:2019rvt}). These proposed ultra-compact object studies are motivated by the hope that at least the echo from one of those objects will eventually be detected in the near future.

Recently, Carballo-Rubio \cite{Carballo-Rubio:2017tlh} proposed a modified TOV equation generated by semi-classical gravity theory.
Solutions obeying boundary conditions are found. They arise from the pressure equation $p'$ with a positive sign and are identified with a nontrivial combination of the black stars and gravastars. This calculation is done by defining a suitably new constant $\lambda>1$ that relates all pressure $p$, mass $m$, and energy density $\rho$, hence the EoS are unique. The form of pressure equation $p'$ with the negative sign has also been analyzed by Ho and Matsuo~\cite{Ho:2017vgi} to show, using constant energy density, that the Buchdahl limit can be violated without making the pressure goes to infinity.

    In this work, we investigate the form of pressure equation $p'$ with negative sign much further. This paper is organized as follows. In section \ref{semi}, we briefly revisit the models (Refs.~\cite{Carballo-Rubio:2017tlh} and~\cite{Ho:2017vgi}). We discuss the numerical scheme in Appendix (page~\pageref{appendix}), which is crucial to justify our results in the following section~\ref{result}. In section~\ref{result} part A, we discuss the model in more detail with non-negative pressure and energy density assumption, and we use a simple linear EoS. We look at the effect from the semi-classical terms compared to the usual TOV equation. In section~\ref{result} part B, we discuss the numerical results. Finally, we summarize our work in Section~\ref{concl}.

	\section{COMPACT STARS IN THE THEORY OF SEMI-CLASSICAL GRAVITY}
	\label{semi}
In ref.~\cite{Carballo-Rubio:2017tlh} a new type of TOV equations can obtained by using the renormalized energy-momentum tensor and solving the modified Einstein field equation (EFE)
	\begin{equation}
	G_{\mu\nu}=\frac{8\pi G}{c^4} T_{\mu\nu} + 96\pi^2 l_p^2 \langle \hat{T}_{\mu\nu} \rangle,\label{eq:start}
	\end{equation}
	where $l_p$ is the so-called \textit{renormalized Planck length}. This quantity is related to the usual Planck length $L_\text{Pl}=\sqrt{\hbar G/c^3}$ by
	\[
	l_p = \sqrt{N\over 12\pi}L_\text{Pl},
	\]
where $N\gg1$ is the number of particle fields. From now on, we shall use the natural units ($c=1$). The renormalized stress-energy tensor (RSET) describes the quantum vacuum polarization of $N\gg1$ matter fields. We shall symbolize the usual Planck length as $L_\text{Pl}$ so that there is no ambiguity with the normalized one $l_p$, which will be treated as an adjustable coupling constant.

    Similar to the derivation of the usual TOV equation, the space-time metric $g_{\mu\nu}$ is the static spherically symmetric metric
	\begin{eqnarray}
	ds^2&=&ds^2_{(2)}+r^2(d\theta^2+\sin^2\theta d\varphi^2),\\
	ds^2_{(2)}&=&-e^{\nu(r)}dt^2+{dr^2\over 1-2Gm(r)/r},
	\end{eqnarray}
	which has a time symmetry denoted mathematically by a timelike Killing vector $\xi=\partial_t$,
	and the energy-momentum tensor can be expressed by the usual perfect fluid
	\begin{equation}
	T^{\mu\nu}=(\rho+p)u^\mu u^\nu+pg^{\mu\nu},
	\end{equation}
	where $u^\mu$ is the 4-velocity of the fluid, which is timelike and normalized ($u^\mu u_\mu=-1$), and $g^{\mu\nu}$ is the inverse of the metric.
	The RSET is given by the so-called {\it s}-wave Polyakov approximation (see Ref.~\cite{Fabbri:2005} page 216)
	\begin{equation}
	\langle \hat{T}_{\mu\nu} \rangle = {\delta^a_\mu \delta^b_\nu\over 4\pi r^2} \langle\hat{T}_{ab}^{(2)}\rangle,
	\end{equation}
	where indices $a$ and $b$ denote 2-dimensional space-time coordinates $ds^2_{(2)}$ where $\langle\hat{T}_{ab}^{(2)}\rangle=\langle 0|\hat{T}_{ab}^{(2)}|0\rangle$ is evaluated. The Boulware state $|0\rangle$ is associated with the Killing vector $\xi$ from the metric, i.e., $\xi^\mu=\delta^\mu_t$ which implies $\xi_\mu=-e^{\nu}\delta_{\mu t}$. It is a usual practice to use null coordinate to obtain $\langle\hat{T}_{ab}^{(2)}\rangle$ and then transform back to $(t,r)$ coordinate. It was shown in~\cite{Barcelo:2011bb} that there is a shortcut equivalent to the usual practice, i.e.,
	\begin{equation}
	\langle\hat{T}_{ab}^{(2)}\rangle={1\over 48\pi}\left( 
	R^{(2)}g_{ab}+A_{ab}-{1\over2}g_{ab}A,
	\right)
	\end{equation}
	where $R^{(2)}$ is the Ricci scalar from the 2-dimensional metric and $A_{ab}$ is related to the Killing vector $\xi$ (in this case $|\xi|=e^{\nu/2}$) by
	\begin{equation}
	A_{ab}={4\over|\xi|}\nabla_a\nabla_b|\xi|.
	\end{equation}
	From Wald's axioms~\cite{Wald:1995yp}, the contracted Bianchi identity $\nabla_\mu \langle \hat{T}^{\mu\nu}\rangle=0$ should be satisfied. This can be checked by substituting the components of the RSET the identity.
	
	Then from the contracted Bianchi identity $\nabla_\mu T^{\mu\nu}=0$ and the above modified EFE \eqref{eq:start} we obtain three equations of motion
	\begin{eqnarray}
	-p'-{(p+\rho)\over 2}{\nu'}&=&0,\label{eq:p}\\
	{\nu'\over r}-{2Gm\over r^3(1-2Gm/r)}&=&{8\pi Gp\over 1-2Gm/r}-{l_P^2\over 4}\left({\nu'\over r}\right)^2,\label{eq:nu'}\\
	{2Gm'\over r^2}&=&8\pi G \rho + {l_P^2\over r^2}\left[
	\left(1-{2Gm\over r}\right)\left(\nu''+\left(\nu'\right)^2\right)\right.\nonumber\\
	&&\left.
	-\left({Gm'\over r}-{Gm\over r^2}\right){\nu'}-{3\over 4}\left(1-{2Gm\over r}\right)\left(\nu'\right)^2\right].
	\label{eq:m}
	\end{eqnarray}
	Here the prime denotes differentiation with respect to $r$.
	
	There are two roots of the metric solution $\nu'$ from Eq.~\eqref{eq:nu'}: 
	\begin{equation}
		\nu'=-{2r\over l_P^2}\left(1\pm\sqrt{1+{l_p^2\over r^2}{2Gm\over r}{(1+4\pi r^3 p/m)\over(1-2Gm/r)}}\right).\label{eq:sign}
	\end{equation}
It leads to two different expressions for both of $p'$ and $m'$. The negative and positive signs on the right-hand side will be named as \emph{negative} and (\emph{positive}) branches. In the limit of $l_p\to 0$, the negative branch approaches the standard GR-TOV equation while the positive branch does not. In the next section, we focus on this negative branch. Regarding the validity of our numerical calculation, we provide the detailed discussions in Appendix in page \pageref{appendix}.

	\section{RESULTS AND DISCUSSION}
	\label{result}
	
	\subsection{Modified TOV Equations from Semi-classical Gravity}

	Here we investigate the model in Ref.~\cite{Carballo-Rubio:2017tlh} with phenomenological point of view, i.e, we only consider non-negative $p$. Since in general $(l_p/r)^2X$ is not constant, where
	\begin{equation}
	X={2Gm\over r}{(1+4\pi r^3 p/m)\over(1-2Gm/r)},
	\end{equation}
	there was a hope for a more general EoS than \eqref{eq:rho}, that is by relaxing the requirement $\lambda=$ constant so that $\lambda$ is a function of $r$. By inspecting equation
	\begin{equation}
	p'={(p+\rho)r\over l_P^2}\left(1+k\sqrt{1+{l_p^2\over r^2} X}\right)
	\end{equation}
	where $k=\pm 1$,
	we discuss it in three cases: (1) ${l_p^2\over r^2} X<1$, (2) ${l_p^2\over r^2} X>1$ and (3) ${l_p^2\over r^2} X=1$. Clearly $X>0$, since $X\leq0$ must be from $2Gm/r\geq1$. From case (1), by binomial expansion with respect to ${l_p^2\over r^2} X$ to only a leading order, we obtain
	\begin{equation}
	p'=\begin{cases}
	{\rho+p\over 2}\left({4r\over l_p^2 X}+{X\over r}\right), & k=+1,\\
	-{\rho+p\over 2}{X\over r}, & k=-1.
	\end{cases}
	\end{equation}
	Since for any reasonable perfect fluid demand positive pressure and surface of the star demand zero pressure, we need $p'<0$, which is not satisfied by $p'$ from $k=+1$ while $k=-1$ is just standard TOV equation. From case (2), again by binomial expansion with respect to ${r^2/(l_p^2 X)}$, this  leads us to
	\begin{equation}
	p'=k {(\rho+p)\sqrt{X}\over l_p}.
	\end{equation}
	This expression eliminates $k=+1$. From the case (3), this leads us to
\begin{equation}
p'={(p+\rho)r\over l_P^2}\left(1+k\sqrt{2}\right),
\end{equation}
which again imply that we should choose $k=-1$.
Hence from now we shall only investigate $p'$ generated from $k=-1$ so that both $p_c>0$ and $p(R)=0$ are satisfied.
	
	Notice that Eq.~\eqref{eq:m} has $m'$ on  both sides, the full expression of $m'$ is rather lengthy.
	Since we choose $k=-1$, we use
	\begin{equation}
	p'={(p+\rho)r\over l_P^2}\left(1-\sqrt{1+{l_p^2\over r^2} X}\right),\label{eq:p'}
	\end{equation}
	then we have
	\begin{equation}
	m'={4\pi \rho r^2} \left({1+\sum_{i=1}^7 A_i\over 1+\sum_{i=1}^4 B_i}\right),\label{eq:m'}
	\end{equation}
	where the nominator consists of
	\begin{eqnarray}
	A_1&=&\frac{3 l_p^2 p}{\rho r^2 \sqrt{1+{l_p^2\over r^2} X}},\\
	A_2&=&-\frac{3 l_p^2 m (1+4\pi r^3 p/m)}{4\pi \rho r^5 \sqrt{1+{l_p^2\over r^2} X}},\\
	A_3&=&-\frac{ G l_p^2 m^2 (1+4\pi r^3 p/m)}{2\pi \rho r^6 (1-2Gm/r) \sqrt{1+{l_p^2\over r^2} X}},\\
	A_4&=&-\frac{ m \left(1-\sqrt{1+{l_p^2\over r^2} X}\right)}{4\pi \rho r^3},\\
	A_5&=&-\frac{ \left(1-2Gm/r\right) \left(1-\sqrt{1+{l_p^2\over r^2} X}\right)}{4\pi G\rho r^2},\label{eq:A5}\\
	A_6&=&\frac{\left(1-2Gm/r\right) \left(1-\sqrt{1+{l_p^2\over r^2} X}\right)^2}{8\pi G\rho l_p^2},\\
	A_7&=&\frac{ l_p^2 }{ \rho r \sqrt{1+{l_p^2\over r^2} X}}p',
	\end{eqnarray}
	and the denominator consists of
	\begin{eqnarray}
	B_1&=&\frac{4 \pi l_p^2 pr}{ m \sqrt{1+{l_p^2\over r^2} X}},\\
	B_2&=&-\frac{ l_p^2 \left(1+4\pi r^3 p/m\right)}{ r^2 \sqrt{1+{l_p^2\over r^2} X}},\label{eq:B2}\\
	B_3&=&-\frac{2 G l_p^2 m \left(1+4\pi r^3 p/m\right)}{ r^3 \left(1-2Gm/r\right) \sqrt{1+{l_p^2\over r^2} X}},\\
	B_4&=&-{\left(1-\sqrt{1+{l_p^2\over r^2} X}\right)}.
	\label{eq:eom-end}
	\end{eqnarray}
	Since $m'$ is complicated, we need to be careful of fixing the constants and the initial data of pressure.

Before solving those equations numerically, let us investigate both $p'$ and $m'$ at limit $l_p\to 0$. 
Expanding $p'$ with respect to $l_p$, we have
\begin{equation}
	p'=-\frac{G \left(m+4 \pi  r^3 p\right) (\rho+p)}{r (r-2 G m)}+\frac{G^2 l_p^2 \left(m+4 \pi  r^3 p\right)^2 (\rho+p)}{2 r^3 (r-2 G m)^2}+\mathcal{O}(l_p^3).
	\label{eq:expandlp1}
\end{equation}
Let us expand these $A_i$ and $B_i$ w.r.t. $l_p$:
\begin{eqnarray}
	A_1&=&\frac{l_p^2 }{r^2 }\frac{3 p}{\rho}
	+\mathcal{O}(l_p^3)>0,\\
	A_2&=&-\frac{l_p^2 }{r^2}\frac{3 m\left(1+4 \pi r^3 p/m\right)}{4 \pi r^3 \rho}
	+\mathcal{O}(l_p^3)<0,\\
	A_3&=&-\frac{l_p^2}{r^2}\frac{G m^2\left(1+4 \pi r^3 p/m\right)}{2 \pi r^4 \rho (1-2 G m/r)}
	+\mathcal{O}(l_p^3)<0,\\
	A_4&=&\frac{l_p^2}{r^2}\frac{G m^2 \left(1+4 \pi r^3 p/m\right)}{4 \pi r^4 \rho ( 1-2 G m/r)}
	+\mathcal{O}(l_p^3)>0,\\
	A_5&=&\frac{l_p^2}{r^2}\frac{ m\left(1+4 \pi r^3 p/m\right)}{4 \pi r^3 \rho }
	+\mathcal{O}(l_p^3)>0,\\
	A_6&=&\frac{l_p^2}{r^2}\frac{Gm^2 \left( 1+4 \pi r^3 p/m\right)^2}{8 \pi r^4 \rho ( 1-2 G m/r)}
	+\mathcal{O}(l_p^3)>0,\\
	A_7&=&-\frac{l_p^2}{r^2}\frac{Gm (1+p/\rho) \left( 1+4 \pi r^3 p/m\right)}{r (1-2G m/r)}
	+\mathcal{O}(l_p^3)<0,
\end{eqnarray}
and
\begin{eqnarray}
	B_1&=&\frac{l_p^2}{r^2}\frac{4 \pi r^3  p}{ m}
	+\mathcal{O}(l_p^3)>0,\\
	B_2&=&-\frac{l_p^2}{r^2} \left(1+\frac{4 \pi r^3 p}{ m}\right)
	+\mathcal{O}(l_p^3)<0,\\
	B_3&=&-\frac{l_p^2}{r^2} \frac{2 G m\left(1+4 \pi r^3 p/m\right)}{r(1-2 G m/r)}
	+\mathcal{O}(l_p^3)<0,\\
	B_4&=&\frac{l_p^2}{r^2} \frac{Gm \left( 1+4 \pi r^3 p/m\right)}{r (1-2 G m/r)}
	+\mathcal{O}(l_p^3)>0,
\end{eqnarray}
Since we know that $(Gm/r),(r^3p/m),(r^3\rho/m)$ and $(p/\rho)$ are dimensionless, notice that no singularity occur on each $A_i$ and $B_i$ as $r=r_c\sim 0$.
Summing them up,
\begin{eqnarray}
	\sum_{i=1}^7 A_i &=& \frac{l_p^2 }{r^2 }\left[ \frac{3 p}{\rho}
	-2\frac{ m\left(1+4 \pi r^3 p/m\right)}{4 \pi r^3 \rho}
	-\frac{G m^2 \left(1+4 \pi r^3 p/m\right)}{4 \pi r^4 \rho ( 1-2 G m/r)}
	\right.\nonumber\\
	&&\left.
	+\frac{Gm^2 \left( 1+4 \pi r^3 p/m\right)^2}{8 \pi r^4 \rho ( 1-2 G m/r)}
	-\frac{Gm (1+p/\rho) \left( 1+4 \pi r^3 p/m\right)}{r (1-2G m/r)}
	\right]
	+\mathcal{O}(l_p^3)
	\label{eq:Ai}
	,\\
	\sum_{i=1}^4 B_i &=&\frac{l_p^2 }{r^2 }\left[
	-1
	-\frac{Gm \left(1+4 \pi r^3 p/m\right)}{r(1-2 G m/r)}
	\right]
	+\mathcal{O}(l_p^3)
	.\label{eq:expandlp1-end}
\end{eqnarray} 

The value of $\rho_0$ plays a significant role. On the surface of the object, $p=0$, $\rho=\rho_0$, $r=R$ and $m=M$, so we have
\begin{equation}
	p'(R)=-\frac{G M \rho _0}{R^2(1-2 G M /R)}+\frac{G^2 l_p^2 M^2 \rho _0}{2 R^5 (1-2 G M/R)^2}+\mathcal{O}(l_p^3)
\end{equation}
and
\begin{eqnarray}
	\sum_{i=1}^7 A_i &=& \frac{l_p^2 }{R^2 }\left[ 
	-2\frac{ M}{4 \pi R^3 \rho_0}
	-\frac{G M^2 }{4 \pi R^4 \rho_0 ( 1-2 G M/R)}
	\right.\nonumber\\
	&&\left.
	+\frac{G M^2 }{8 \pi R^4 \rho_0 ( 1-2 G M/R)}
	-\frac{G M }{R (1-2G M/R)}
	\right]
	+\mathcal{O}(l_p^3)
	,\\
	\sum_{i=1}^4 B_i &=&\frac{l_p^2 }{R^2 }\left[
	-1
	-\frac{G M }{R(1-2 G M/R)}
	\right]
	+\mathcal{O}(l_p^3)
	,
\end{eqnarray} 
hence $m'(R)$ finite since $\rho_0\neq0$. It seems that if the value of $\rho_0$ goes to zero, the changes of both $R$ and $M$ might be significant.

Now in the center $r=r_c\sim 0$ we have
\begin{eqnarray}
	\sum_{i=1}^7 A_i &=& \frac{l_p^2 }{r_c^2 }\left[ \frac{3 p_c}{\rho_c}
	-2\frac{\left(1+3p_c/\rho_c\right)}{3}
	-\frac{G m_c^2 \left(1+3p_c/\rho_c\right)}{4 \pi r_c^4 \rho_c ( 1-2 G m_c/r_c)}
	\right.\nonumber\\
	&&\left.
	+\frac{G m_c^2 \left( 1+3p_c/\rho_c\right)^2}{8 \pi r_c^4 \rho_c ( 1-2 G m_c/r_c)}
	-\frac{G m_c (1+p_c/\rho_c) \left( 1+3p_c/\rho_c\right)}{r_c (1-2G m_c/r_c)}
	\right]
	+\mathcal{O}(l_p^3)
	\nonumber\\
	&=&\frac{l_p^2 }{r_c^2 }\left[\frac{p_c}{\rho_c}-\frac{2}{3}\right]+\mathcal{O}(r_c^2)
	\label{eq:Ac}
	,\\
	\sum_{i=1}^4 B_i &=&\frac{l_p^2 }{r_c^2 }\left[
	-1
	-\frac{G m_c \left(1+3p_c/\rho_c\right)}{r_c(1-2 G m_c/r_c)}
	\right]
	+\mathcal{O}(l_p^3)
	= -\frac{l_p^2 }{r_c^2 }+\mathcal{O}(r_c^2)
	,\label{eq:Bc}
\end{eqnarray} 
where we have substitute the mass in this limit $m_c=(4/3)\pi\rho_cr_c^3$.
Then we have
\begin{equation}
p'(r_c)\sim-\frac{4}{3} \pi  G r_c (\rho_c+p_c) (\rho_c+3 p_c)+\frac{8}{9} \pi ^2 G^2 l_p^2 r_c (\rho_c+p_c) (\rho_c+3 p_c)^2
\label{eq:approxprc}
\end{equation}
and
\begin{equation}
	m'(r_c)
	\sim 4\pi r_c^2\rho_c\left(
	\frac{1+(l_p/r_c)^2(3p_c/\rho_c-2/3)}{1-(l_p/r_c)^2}
	\right).
	\label{eq:approxmrc}
\end{equation}
        These equations are valid only when $r$ near $r_c$ and clearly $m'(r_c)$ is singular when $r_c=l_p$.

Suppose that we have at the center $r_c=\alpha l_p$ where $\alpha>0$. Then this means that $\alpha\neq 1$. Notice that Eq.~\eqref{eq:approxprc} can be rewritten as
\begin{equation}
	p'(r_c)\simeq-\frac{4}{3} \pi  G \rho_c ^2 \left(1+\frac{p_c}{\rho_c}\right) \left(1+3 \frac{p_c}{\rho_c}\right)\alpha l_p
	\left[1-\frac{2}{3} \pi  G  \rho_c  \left(1+3 \frac{p_c}{\rho_c}\right) l_p^2\right].
	\label{eq:approxprc2}
\end{equation}
The second term in the square bracket can be bigger than the first term if $\alpha l_p^3$ is large enough than $\alpha l_p$. Hence, we conclude that $\alpha l_p^3$ should be sufficiently small such that $p'(r)<0$. Looking at Eq.~\eqref{eq:approxprc2}, then the higher-order terms should have the pattern $\mathcal{O}\left(\alpha l_p^{2n+1}\right)$, with $n$ an integer. This means that we should hold $l_p$ first then adjusting $\alpha$ such that the second term in Eq.~\eqref{eq:approxprc2} does not dominate the first one, i.e.,
\begin{equation}
l_p<\sqrt{\frac{3}{2\pi G (\rho_c+3p_c)}}
\text{, or equivalently, }
N<\frac{18}{L_\text{Pl}^2 G (\rho_c+3p_c)}
.\label{eq:lpupperbound}
\end{equation}
  Hence the upper bound of $l_p$ depends on both $p_c$ and the EoS, which should satisfy the strong energy condition.

\textcolor{black}{By defining an arbitrary positive valued constant $\alpha=r_c/l_p$, Eq. \eqref{eq:approxmrc} becomes
	\begin{equation}
	m'(r_c)
	\sim 4\pi r_c^2\rho_c\left[1-
	\frac{3p_c/\rho_c+1/3}{1-\alpha^2}
	\right].
	\end{equation}
	Logically the mass should grow from center to the surface, so $m'(r)>0$ ($0<r\leq R$).
	Notice that if $\alpha>1$ then $m'(r_c)>0$ is trivially satisfied. On the other hand, if $\alpha<1$ then $m'(r_c)>0$ can happen if the second term inside the square bracket is less than unity. This implies
	\begin{equation}
	3p_c<(2/3-\alpha^2)\rho_c.
	\end{equation}
	If $\alpha=\sqrt{2/3}$ then $p_c<0$. This negative center pressure is contradicting with our previous assumption that $p>0$ and $\rho>0$ inside the star.
	If $\alpha>\sqrt{2/3}$ then
	\begin{equation}
	\rho_c<-\left({3p_c\over \alpha^2-2/3}\right)<0.
	\end{equation}
	This expression is also a contradiction with our previous assumption. If $\alpha<\sqrt{2/3}$ then
	\begin{equation}
	\rho_c>\left({3p_c\over 2/3-\alpha^2}\right)>4.5p_c,
	\end{equation}
	which imply speed of sound squared $dp/d\rho=w<2/9$, lower than the upper bound from QCD and causality, where $w\leq 1/3$ and $w\leq 1$, respectively~\cite{Moustakidis:2016sab}. According to Ref.~\cite{Urbano:2018nrs}, the maximum compactness produced by a linear equation of state (EoS) $\rho=p/w+\rho_0$ is
	\begin{equation}
	\left({2GM\over R}\right)_\text{max}\sim\frac{8}{9 \left(\frac{0.51 w +0.77}{w  (w +4.18)}+1\right)},
	\end{equation}
	which is a monotonically increasing function of $w$ as $w$ grow but cannot go beyond the Buchdahl limit $2GM/R=8/9$. So $w<2/9$ will produce less compact stars than $w=1/3$. Also notice that this condition $m'(r)$ applies only at $r=r_c$, so it may not be true for $r>r_c$.}
	
	\textcolor{black}{
	The case of $\alpha<1$ may still be investigated with negative valued $m_c$, which is possible according to Ref~\cite{Anastopoulos:2020mrt}. From Eq.~\eqref{eq:expandlp1-end} then we can see that singularity of $m'(r)$ may not be there anymore for $\alpha=1$. If we demand $\sum_i B_i>0$ then 
	\begin{equation}
	m_c<\frac{\rho_c r_c}{G \rho_c-3 G p_c}\equiv m_{c,max} \label{eq:mcmax}
	\end{equation}
	and since $m_c<0$ we obtain another restriction for the EoS
	\begin{equation}
	w>1/3.
	\end{equation}
	Notice that $m_{c,max}$ can be made close to zero so we can use limit $m_c\to 0^-$. Eq.~\eqref{eq:expandlp1} in the limit of $m_c\to 0^-$ becomes $p'(r_c)<0$ so $p$ will decrease from $p_c$. In this limit, Eq.~\eqref{eq:Ai} becomes
	\begin{equation}
	\sum_i A_i\sim \alpha^{-2}\left[{p_c\over \rho_c}-{2\over 3}\right].
	\end{equation}
	Since we demand $\sum_i B_i>0$ and $m'(r)>0$, then
	\begin{equation}
	w>2/3.
	\end{equation}
	This is from considering $m_c\to 0^-$. If on the other hand we consider $m_c\to-\infty$, notice that Eq.~\eqref{eq:expandlp1} set the minimum value of $m_c$, i.e.,
	\begin{equation}
	0>m_c>-4\pi r_c^3 p_c\equiv m_{c,min}. \label{eq:mcmin}
	\end{equation} 
	Notice that $m_{c,min}$ can be very close to zero so the range of $m_c$ is quite small. Thus for the case of $m_c<0$, we obtain that both $w$ and $m_c$ are restricted to $w>2/3$ and $m_{c,min}<m_c<m_{c,max}$. Due to the tight restriction of $m_c<0$, we do not focus on $m_c<0$ case in the following sections but rather on the usual positive valued $m_c=4\pi r_c^3\rho_c/3$.
	}
	
\begin{figure}
	\centering
	\includegraphics[width=0.99\linewidth]{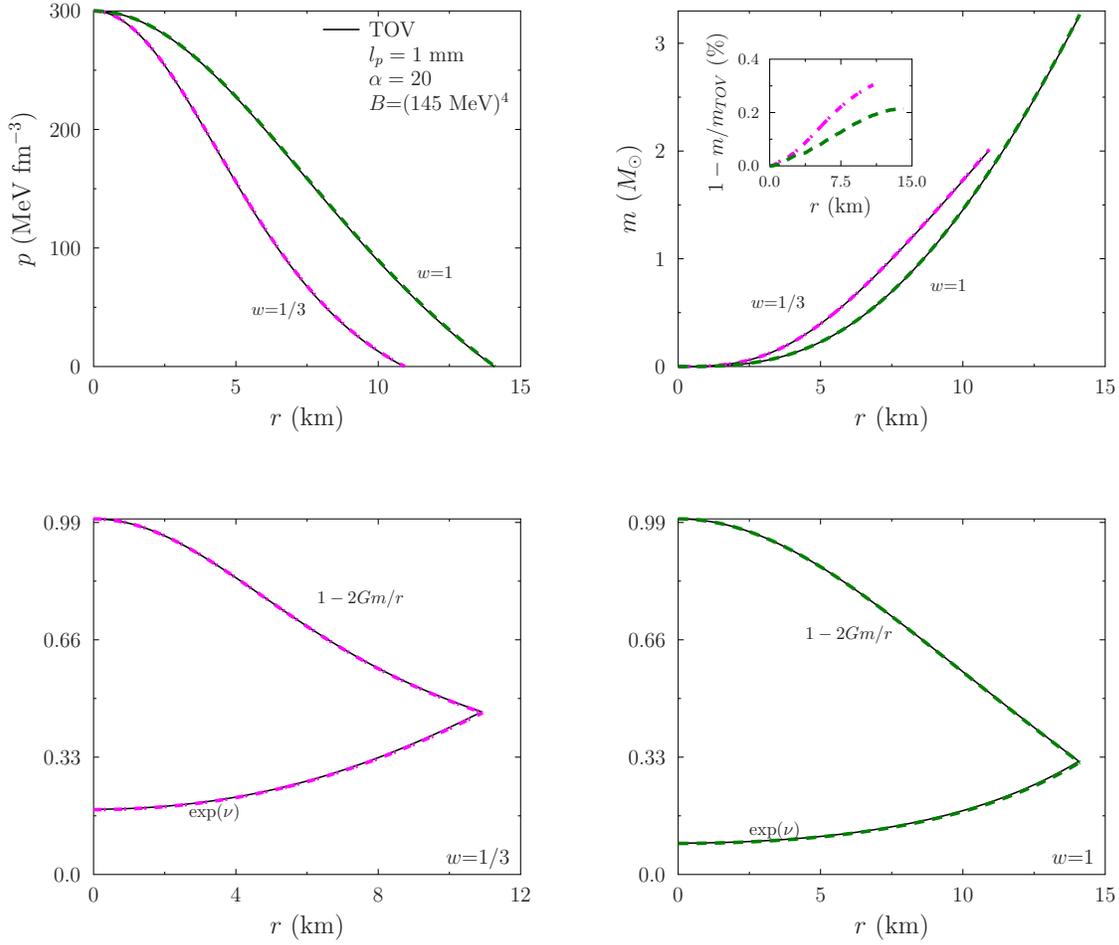}
	\caption{Here we have profiles of $p$ and $m$ in the upper panels from $l_p=1$ mm and $\alpha=20$ and we vary the speed of sound squared $w$. The metric function $\exp(\nu)$ and $1-2Gm/r$ are also shown for each $w$ in the lower panels.}
	\label{fig:plotprofil_varw}	
\end{figure}
\begin{figure}
	\centering
	\includegraphics[width=0.5\linewidth]{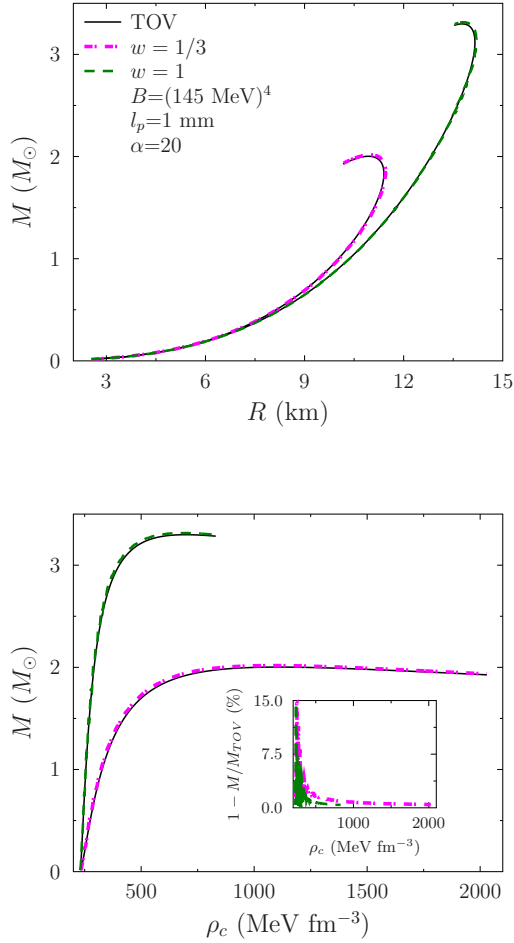}
	\caption{Here we show the M-R curve from varying $w$. The semi-classical correction does not produce significant discrepancy compared to TOV from GR.}
	\label{fig:plotmrvarw}
\end{figure}
\begin{figure}
	\centering
	\includegraphics[width=0.5\linewidth]{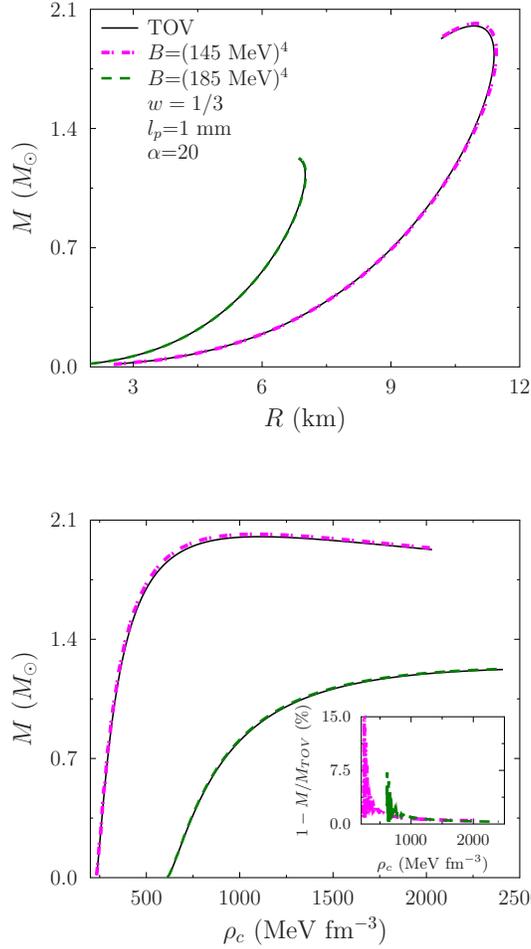}
	\caption{Here we show the M-R curve from varying $B$. Decreasing $B$ produces higher mass and larger radius.}
	\label{fig:plotmrvarB}
\end{figure}
\begin{figure}
\centering
\includegraphics[width=0.5\linewidth]{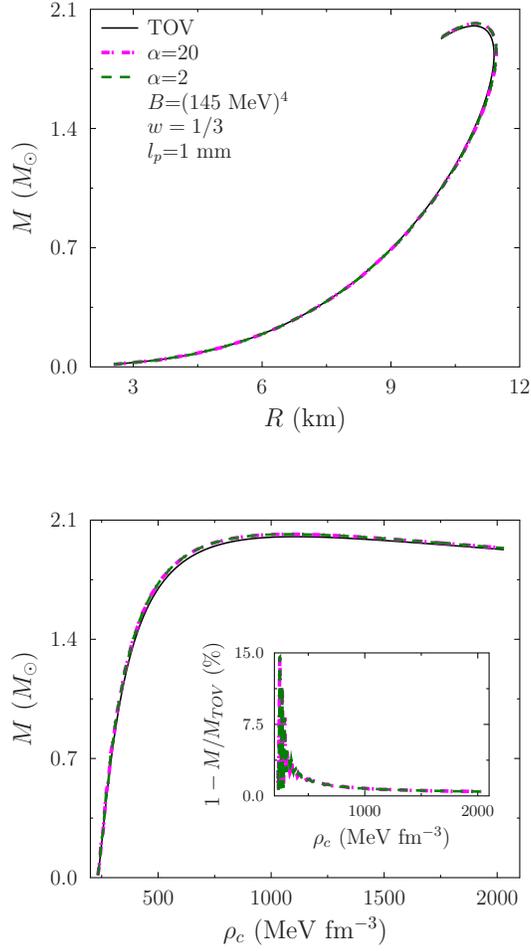}
\caption{Here we show the M-R curve from varying $\alpha$. The choice of $\alpha$ does not show any difference in the result.}
\label{fig:plotmrvaralpha}
\end{figure}
\begin{figure}
\centering
\includegraphics[width=0.5\linewidth]{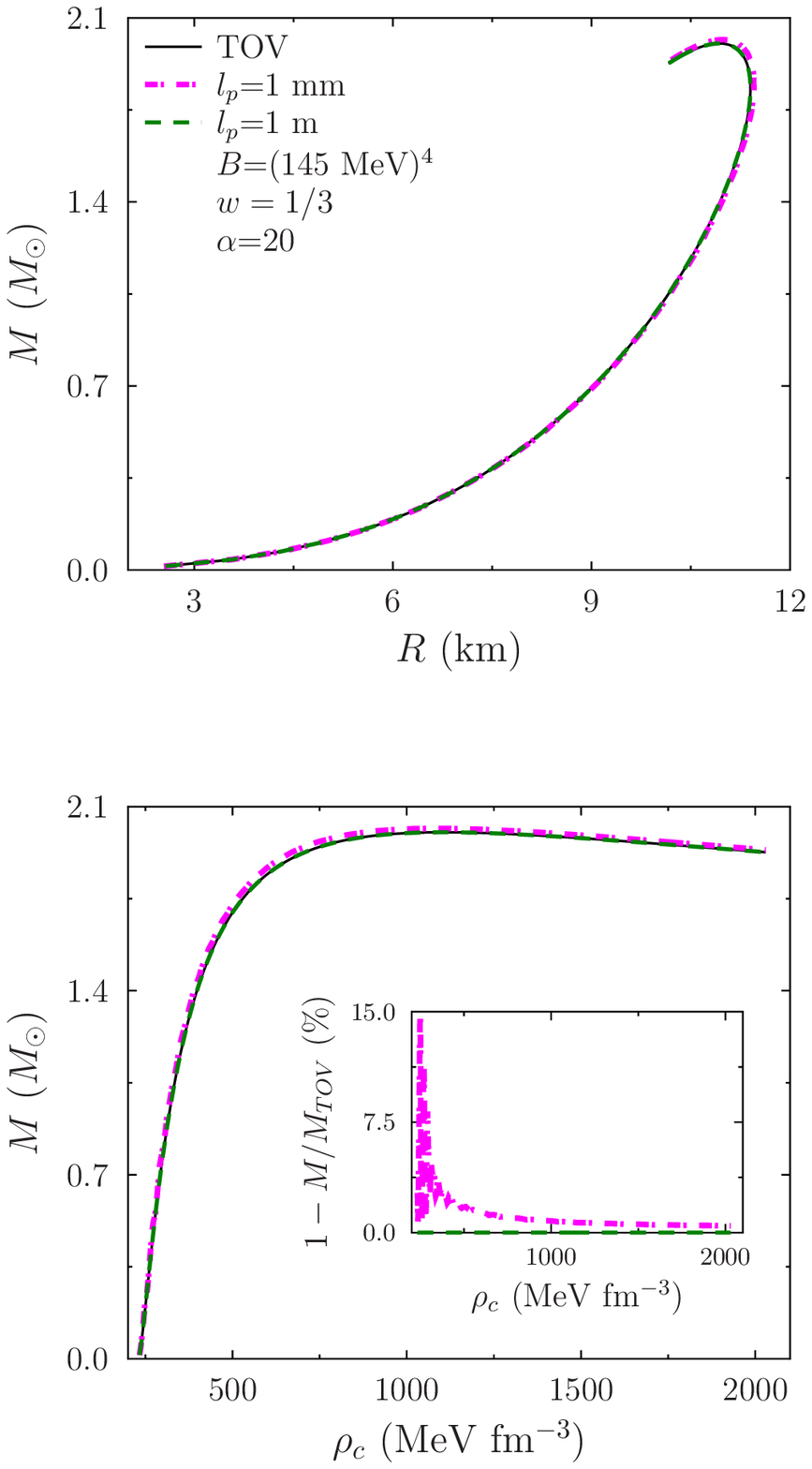}
\caption{Here we show the M-R curve from varying $l_p$. Increasing $l_p$ also does not significantly shift the M-R curve.}
\label{fig:plotmrvarlp}
\end{figure}

\subsection{Numerical Results}

 The EoS used here have the following linear form
\begin{equation}
\rho(p)=p/w+\rho_0,\label{eq:lineos}
\end{equation}
where $w$ is the speed of sound squared from thermodynamics and $\rho_0$ is a positive valued constant. The constant $\rho_0$ cannot be zero since it implies no solution and cannot be negative, or it will violate weak energy condition~\cite{Urbano:2018nrs}. The constant $w$ is restricted to at least two conditions~\cite{Moustakidis:2016sab}: (1) from causality, we have $0<w\leq1$ and (2) from QCD and other known theories, we have $0<w\leq1/3$. \textcolor{black}{Here we focus on compact stars with largest compactness possible, so we shall choose $1/3\leq w\leq1$ and $\alpha>1$.}

To integrate both $p'$ and $m'$ numerically we need to fix $l_p, \alpha, \rho_0$, and $p_c$ carefully such that all boundary conditions are satisfied. Here we shall use natural units for those four constants following Ref.~\cite{Diener:2008bj}. In this set of units called ``NS'' units, $r$ is in metres, both $p$ and $\rho$ are in MeV/fm$^3$ and $m$ is in MeV m$^3$/fm$^3$. 
The compactness is defined as $C=GM/R$. We use $\rho_0=4B$, where $B$ is the so-called bag constant from the MIT bag model~\cite{Mannarelli:2018pjb}. 

The upper bound of $l_p$ can now be determined. It is usual to have both $p_c$ and $\rho_0$ at most $\sim 10^{3}$ MeV/fm$^3$, then by Eq.~\eqref{eq:lpupperbound}, we have $l_p\lesssim 10$ km. Since the observed neutron stars have $R\sim 10$ km, we have $l_p<R$, which is trivial if we consider $l_p$ to be related to the Planck length. This finding is equivalent to fixing the upper bound for the dimensionless parameter $N$, which is $N<10^{79}$. This number is a huge quantity since the RSET should satisfy (in SI units using $\hbar c=3.162 \times 10^{-27}$ kg m$^3$ s$^{-2}$)   	
\begin{equation}
|\langle T_{\mu\nu} \rangle| < \frac{|T_{\mu\nu}|}{3.162 \times 10^{52} \text{ kg m$^3$ s$^{-2}$}}.
\end{equation}

Since the smallness of $l_p$ often make calculators cannot detect the second term in the square root in Eq. \eqref{eq:p'}, this form tends to make $p'=0$ in some $r$. To evade this, we use an equivalent form of Eq.~\eqref{eq:p'}, i.e.,
\begin{equation}
p'=-\frac{(p+\rho)X}{r\left(1+\sqrt{1+(l_p/r)^2 X}\right)}.
\end{equation}

The following numerical results show that the contribution of RSET does not significantly affect the maximum mass. An example for profiles from varying $w$ are shown in Fig.~\ref{fig:plotprofil_varw}.  We see that the contribution of $w$ does make higher mass and larger radius resulting in an ultra-compact star, but we will not discuss its echo property since this compactness can also be obtained from the standard TOV equation in GR.

Varying $B$, $\alpha$ and $l_p$ lead to similar profiles. The M-R curves from varying $w$, $B$, $\alpha$ and $l_p$ are shown in Figs.~\ref{fig:plotmrvarw},~\ref{fig:plotmrvarB},~\ref{fig:plotmrvaralpha} and~\ref{fig:plotmrvarlp}, respectively. Here the semi-classical correction does not produce significant discrepancy from arbitrary choice of $p_c$ compared to TOV from GR.

We also had tried using $\alpha<1$, but we cannot find the numerical solution. The mass always goes down to a negative value, even though it increases a little in the beginning. It is interesting to note that for the similar case considered in Ref.~\cite{Ho:2017vgi} their integration method breaks down at $r=l_p$. They use constant energy density, and they integrate the equations from the surface to the core.

	\section{Conclusions}
	\label{concl}	

In this work, we analyze the semi-classical theory of gravity proposed in Ref.~\cite{Carballo-Rubio:2017tlh}. Note that the theory has two different sets of equations characterized by the sign in $p'$. This happened since the Einstein equation for metric solution $\exp(\nu)$ in \eqref{eq:sign} has two roots and $p'$ is related to $\nu'$ by Eq. \eqref{eq:p}. 

From the positive branch, defining a constant $\lambda$ is done to obtain analytic solutions. By definition of $\lambda$, $p$ is related to energy density $\rho$ and mass $m$. The solution's character $p$ is as follows: starting from the negative value of $p$ at $r\to 0$ and increase the $p$  to zero. This solution resembles a combination of the known gravastars or black star models. We show that the numerical solution can be obtained by integrating from its surface to its core.

    From the negative branch, which goes to the usual TOV equation at the formal limit $l_p\to0$, we had reproduced the results from~\cite{Ho:2017vgi} for the case of constant energy density. In this paper, we analyze this negative branch further by applying linear EoS $\rho=p/w+\rho_0$ with $w$=1/3 and  $w$=1, respectively. 
    
    The range of $l_p$ is dependent on the choice of $r_c$. The reason is as follows. Since the equation $m'(r)$ is much more complicated than usual TOV equation (see Eqs.\eqref{eq:m'}-\eqref{eq:eom-end}), it has terms that can make $m'(r)<0$, which may imply negative $m$ since at the center $m_c\sim 0$. These terms are dependent on the values of both $l_p$ and $r_c$ such that $m'(r_c)$ is singular when $l_p=r_c$.
    
    To investigate $m'(r)$ around this singularity, we fix a relation $r_c=\alpha l_p$ with $\alpha\neq1$. We also demand $m'(r)>0$ for $0<r<R$ since it is usual practice to expect the mass increase from the center to surface. We found that, with some approximation steps, $0<\alpha<\sqrt{2/3}$ demands a very restricted set of EoS with a speed of sound squared $w<2/9$. This value implies the compactness lowered, so we are not discussing this aspect. For $\sqrt{2/3}\leq\alpha<1$, this demands either $p_c<0$ or $\rho_c<0$ which contradict our starting assumption that both pressure and energy density should not be negative. For the case of $\alpha>1$, we can use any EoS. Thus our analysis agrees with Ref.~\cite{Ho:2017vgi}, i.e., their integration breaks down at $r=l_p$.
    
    The value of $l_p$ has an upper bound-constrained by the pressure at the center $p_c$. From our approximation, we found from the dimensional analysis that $l_p<10$ km, which is of order $R$. This upper bound implies $N<10^{79}$. This fact implies that $l_p$ can be much larger than Planck length. 
    
    Through varying all four parameters $w$, $B$, $\alpha$, and $l_p$, our numerical results show that the M-R curves are indistinguishable compared to the TOV equation results in GR. Hence the parameter $l_p$, though not in the limit $l_p\to L_\text{Pl}$, has no significant signature compared to the standard TOV equation. 
    
    Compactness reaching the ultra-compact range can be achieved by adjusting both $w$ and $B$ in the TOV equation system with and without semi-classical correction. Moreover, the authors in Ref.~\cite{Ho:2017vgi} showed that the Buchdahl limit could be violated using constant energy density, which means setting $w$ equal to infinity. Hence, it seems that we need to reach somehow much higher compactness larger than the Buchdahl limit to see the significance of the semi-classical correction. This way might be done either by adjusting decreasing $B$ to near zero or increasing $w$ such that $w>1$, which violates causality. 
    
    Hence we conclude this section with the following points. 
    (1) The structure of the Eqs.~\eqref{eq:p'}-\eqref{eq:eom-end} restricts the range of $\alpha$ and $w$ in some ways that turn out to give results not significant if compared to the usual TOV equation system.  Therefore, similar to the conclusion in Ref.~\cite{Urbano:2018nrs} with GR theory that the ultra-compact objects from the negative branch of semi-classical gravity with a linear equation of state are unable to generate demanding gravitational echoes.
    (2) Our numerical results are consistent with both Ref.~\cite{Urbano:2018nrs} and Ref.~\cite{Ho:2017vgi}. See the detailed discussion in Appendix.
    (3) The positive branch is still open for more detailed numerical analysis. We expect that the quantum effect within semi-classical approximation is only significant for exotic compact objects, considered the solution related to the positive branch.

	\begin{acknowledgments}
	
This work is funded by Publikasi Terindeks Internasional (PUTI) Doktor 2020, No. NKB-614/UN2.RST/HKP.05.00/2020.

	\end{acknowledgments}


	\appendix* 
	
	\section{Discussions Regarding Numerical Method}\label{appendix}
	
In this appendix, we discuss the integration schemes and their results. As a disclaimer, in this paper, all the calculations are done using \textit{Mathematica 10.0}. The schemes are named as \emph{forward} (\emph{backward}, resp.) integration, corresponding to integrating from center to surface (surface to center). It is known that for the standard TOV equation, we can do both forward and backward integration. Here we discuss both forward, and backward integration to obtain numerical solutions for (1) negative branch obtained previously by Ho and Matsuo~\cite{Ho:2017vgi} in the case of constant energy density $\rho$ and (2) the negative branch obtained previously by Carballo-Rubio~\cite{Carballo-Rubio:2017tlh} using constant $\lambda$ trick.
	
	We emphasize these forward and backward integration schemes because we use forward integration in the next section, but both~\cite{Carballo-Rubio:2017tlh} and~\cite{Ho:2017vgi} imply backward integration. We show that for the negative branch, we get consistent results from both backward and forward integration. This evidence justifies our numerical method for the negative branch system that we discussed in the section III onwards.
	
	\subsubsection{Forward and backward integration solutions from negative branch}
	
	Suppose we redefine the $rr$ component of the metric by
	\begin{eqnarray}
	F(r)&=&\sqrt{C(r)\left(1-{2Gm(r)\over r}\right)},\\
	C(r)&=&e^{\nu(r)},
	\end{eqnarray}
	and the constants by $\alpha=l_p^2$ and $\kappa=8\pi G$.
	Then the equations \eqref{eq:nu'} and \eqref{eq:m} are equivalent to equations (5.7) and (5.8) in~\cite{Ho:2017vgi},
	\begin{eqnarray}
	0&=&-\frac{1}{8 r^2 C(r)^2}\left[
	-3 \alpha  F(r)^2 C'(r)^2+4 r C(r)^2 F(r) F'(r)
	\right.\nonumber\\&&\left.
	+2 C(r) F(r) \left(\alpha  C'(r) F'(r)+F(r) \left(\alpha  C''(r)-2 r C'(r)\right)\right)
	\right.\nonumber\\&&\left.
	+2 \kappa  r^2 C(r)^3 (p+\rho)
	\right],\\
	0&=&\frac{1}{4 r^2 C(r)^2}\left[
	\alpha  F(r)^2 C'(r)^2-\alpha  C(r) F(r) \left(F(r) C''(r)+C'(r) F'(r)\right)
	\right.\nonumber\\&&\left.
	-2 C(r)^2 F(r) \left(r F'(r)+F(r)\right)+C(r)^3 \left(2-\kappa  r^2 (\rho-p)\right)
	\right]
	\end{eqnarray} 
To avoid confusion, we keep our symbol for energy density and metric function as $\rho$ and $\nu$, respectively, while in~\cite{Ho:2017vgi} the authors use $m$ and $\rho$. In this paper, we assign $m$ as the mass function. Setting the energy density as constant $\rho=\rho_0$, we can see from \eqref{eq:p} that the pressure becomes
	\begin{equation}
	p(r)=-\rho_0+p_0 e^{-\nu(r)/2}.\label{eq:rhoconst}
	\end{equation}
Here $p_0$ is a constant that should satisfy the boundary condition at the surface $p(R)=0$, i.e., $p_0=\rho_0 \sqrt{1-2GM/R}$ with $M=m(R)$.
	
	Only in this subsection, the integrations are done for the tortoise coordinate $r_*$ which is defined as
	\begin{equation}
	r_*=\int{dr\over F(r)}.
	\end{equation}
	This then give us the equations for $r(r_*)$ and $\nu(r_*)$ as
	\begin{eqnarray}
	0&=&-\frac{1}{8 r(r_*)^2 }\left[
	2\left\{
	\alpha\nu''(r_*)-2r(r_*)\nu'(r_*) r'(r_*)
	\right\}
	\right.\nonumber\\&&\left.
	+4r(r_*)r''(r_*)+2\kappa r(r_*)^2 p_0 e^{\nu(r_*)/2}-\alpha [\nu'(r_*)]^2
	\right],\label{eq:HoMatsuo0}\\
	0&=&\frac{1}{4 r(r_*)^2}\left[
	-\alpha \nu''(r_*)-2(r(r_*) r''(r_*)+[r'(r_*)]^2)
	\right.\nonumber\\&&\left.
	+ e^{\nu(r_*)} [2-\kappa r(r_*)^2(2\rho_0-p_0 e^{-\nu(r_*)/2})]
	\right].\label{eq:HoMatsuo1}
	\end{eqnarray}
	From substracting \eqref{eq:HoMatsuo0} with \eqref{eq:HoMatsuo1} we can see there are two roots of $\eta'(r_*)$, corresponding to the negative and positive branch we had discussed before. Since in this subsection we focus on the negative branch, which goes to TOV equation in the limit $\alpha\to 0$
	\begin{equation}
	\nu '(r_*)=-\frac{2}{\alpha }\left(r(r_*) r'(r_*)-\sqrt{r(r_*)^2 \left(\alpha  \kappa  p_0 e^{\frac{\nu (r_*)}{2}}+r'(r_*)^2-\alpha  \kappa  \rho_0 e^{\nu (r_*)}\right)+\alpha  \left(e^{\nu (r_*)}-r'(r_*)^2\right)}\right). \label{eq:HoMatsuoEq1}
	\end{equation}
	This corresponds to the second equation
	\begin{eqnarray}
	r''(r_*)&=& \frac{1}{4 \left(\alpha -r(r_*)^2\right) r'(r_*)}\left[
	\kappa  r(r_*)^2 e^{\frac{\nu (r_*)}{2}} \left(2 \rho_0 e^{\frac{\nu (r_*)}{2}}-p_0\right) 
	\right.\nonumber\\&&\left.
	\times\left(2 \sqrt{r(r_*)^2 \left(\alpha  \kappa  p_0 e^{\frac{\nu (r_*)}{2}}+r'(r_*)^2-\alpha  \kappa  \rho_0 e^{\nu (r_*)}\right)+\alpha  \left(e^{\nu (r_*)}-r'(r_*)^2\right)}-\alpha  \nu '(r_*)\right)
	\right.\nonumber\\&&\left.
	-\left(2 \sqrt{r(r_*)^2 \left(\alpha  \kappa  p_0 e^{\frac{\nu (r_*)}{2}}+r'(r_*)^2-\alpha  \kappa  \rho_0 e^{\nu (r_*)}\right)+\alpha  \left(e^{\nu (r_*)}-r'(r_*)^2\right)}-\alpha  \nu '(r_*)\right)
	\right.\nonumber\\&&\left.
	\times 2 e^{\nu (r_*)} +4 r(r_*) r'(r_*) \left(\alpha  \kappa  p_0 e^{\frac{\nu (r_*)}{2}}+r'(r_*)^2-\alpha  \kappa  \rho_0 e^{\nu (r_*)}\right)
	\right] \label{eq:HoMatsuoEq2}
	\end{eqnarray}
  So we have a first order equation for $\nu(r_*)$ and second order equation for $r(r_*)$. These depends on three parameters $M$, $R_*=r_*(R)$, $\rho_0$ and two coupling constants $\alpha$ and $\kappa$. The initial conditions at $r_*=R_*$ are
	\begin{eqnarray}
	R_*&=&R+2GM \ln (R/2GM -1),\\
	\nu(R_*)&=&\ln (\sqrt{1-2GM/R}),\\
	r(R_*)&=&R,\\
	r'(R_*)&=&1-2GM/R.
	\end{eqnarray}
Then we do the backward integration. To fix the three input parameters $R, M,$ and $\rho_0$, we follow recipe in~\cite{Ho:2017vgi}: fix $M$ first, then fix $R$ such that $R>2GM$, then explore the value of $\rho_0$. 
	
	To every value of $\rho_0$, we have different shooting parameter values $k$ related to $R$ by $R=2GM/k$, so it is restricted to $0<k<1$. It is also carefully chosen so that at the center the slope of $p$ is not steep ($|p'(r_*\to-\infty)|<\infty$). This result has two reasons. 
	First, assigning $p'(r_*\to-\infty)\sim 0$ is done so that when we integrate the equation in the opposite direction (forward), we can get the same curves. 
	Second, the negative branch goes to the TOV equation in the small $\alpha$ limit and it is known that, in the TOV system, $p'(r=0)\sim 0$ so the negative branch should also has this property for sufficiently small $\alpha$. We assume that $r_*$ is linear to $r$ in the region near $r=0$ so that $p'(r_*\to-\infty)\sim 0$ imply $|p'(r\to 0)|<\infty$.
	
	\begin{figure}
		\centering
		\includegraphics[width=0.5\linewidth]{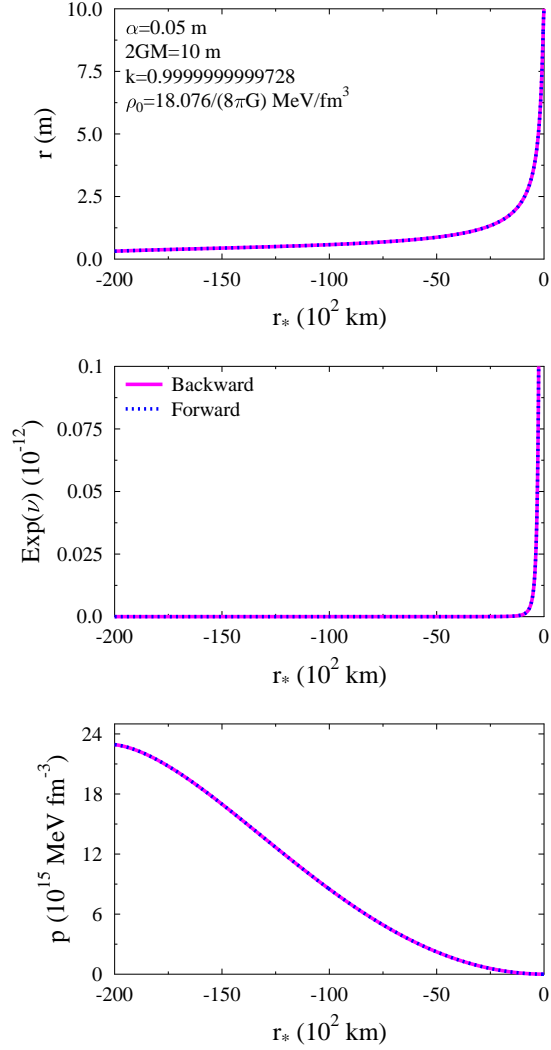}
		\caption{The backward and forward integration on the negative branch gives same numerical curves. Here we integrate them with respect to tortoise coordinate $r_*$. Here $k$ is the shooting parameter to determine $R$ by $R=2GM/k$ so that $p'(r_*=0)\sim 0$. Here $e^{(\nu)}$ goes really close to zero as $r_*\to-\infty$ but never reach zero.}
		\label{fig:plotprofilhomatsuo}
	\end{figure}
	The results from doing backward integration and calculating again using forward integration are shown in Fig.~\ref{fig:plotprofilhomatsuo}. We obtain that the forward integration produces the same curves as the backward one. The pressure, in the region $r_*\to-\infty$, is indeed does not go to Planck scale for the macroscopic size of $R$.

	\subsubsection{Forward and backward integration solutions from positive branch}
	
The author in Ref.~\cite{Carballo-Rubio:2017tlh} had shown that the positive branch have exact solutions whose properties are very similar to a mixture of gravastars and black stars, i.e., the negative pressure in the interior
	\begin{equation}
	p={-1+\mathcal{O}(l_p^2/r^2)\over 8\pi G r^2 R^2}\left[R^2-r^2e^{\frac{(\lambda +1) (r^2-R^2)}{l_p^2}}\right],
	\label{eq:exactp}
	\end{equation}
	with $\lambda>1$ a constant,
	but its energy density is still positive valued
	\begin{equation}
	\rho(r)={1+\mathcal{O}(l_p^2/r^2)\over 8\pi G r^2 R^2}\left[R^2+r^2e^{\frac{(\lambda +1) (r^2-R^2)}{l_p^2}}\right]. \label{eq:exactrho}
	\end{equation}
			Notice that $p(0)<0$ and $\rho(0)>0$.
	Moreover, both satisfy weak, dominant, and null energy conditions. The strong energy conditions are assumed unnecessary since the existence of Casimir energy in experiments violate it. The key point of the derivation of Eqs.~\eqref{eq:exactp} and \eqref{eq:exactrho} is the introduction of a constant $\lambda>1$ that is defined as
	\begin{equation}
	\lambda\equiv\sqrt{1+{l_p^2\over r^2}{2Gm\over r}{(1+4\pi r^3 p/m)\over(1-2Gm/r)}},
	\end{equation}
	where it implies a fixed form of EoS and mass profile
	\begin{eqnarray}
	\rho&=& -p+\frac{l_p^2}{(\lambda +1) r}p',\label{eq:rho}\\
	m&=&\frac{r^3 \left(-8 \pi G l_p^2 p+\lambda ^2-1\right)}{2 G \left(l_p^2+\left(\lambda ^2-1\right) r^2\right)}.
	\label{eq:mlam}
	\end{eqnarray}
	
	Since $p=0$ at $r=R$, we have the compactness determined by $R, l_p$ and $\lambda$ as
	\begin{equation}
	2C={2GM\over R}=\frac{1}{ \left(l_p^2/\left[R^2\left(\lambda ^2-1\right)\right]+1\right)},\label{eq:comp0}
	\end{equation}
	with $M=m(R)$.
	The resulting exact solution for metric is
	\begin{equation}
	\nu(r)=\nu(R)+{\frac{(\lambda +1) (R^2-r^2)}{l_p^2}},\\
	\end{equation}
	with $\nu(R)=\ln\left(1-2GM/R+\mathcal{O}(l_p^2/R^2)\right)$.  The solutions can lead to arbitrary compactness depending on the value of $\lambda$ and are 
	stable by curvature and boundary conditions arguments. 
	
	After some algebra, one can arrive at the following equation of motion
	\begin{eqnarray}
	p'(r)&=&g(r)+p(r) h(r),\label{eq:steepP}
	\\
	g(r)&=&\frac{(\lambda +1)^2 \left(l_p^4+(\lambda -2) l_p^2 r^2-\left(\lambda ^2-1\right) r^4\right)}{4 \pi  G l_p^2 r \left(l_p^2-(\lambda +1) r^2\right) \left(l_p^2+\left(\lambda ^2-1\right) r^2\right)},
	\\
	h(r)&=&\frac{2 (\lambda +1) r \left((2 \lambda +1) l_p^4+\left(\lambda ^3+\lambda ^2-2 \lambda -2\right) l_p^2 r^2-(\lambda -1) (\lambda +1)^2 r^4\right)}{l_p^2 \left(l_p^2-(\lambda +1) r^2\right) \left(l_p^2+\left(\lambda ^2-1\right) r^2\right)}.
	\end{eqnarray}
	We integrate by backward integration starting from $r=R$ and $p(R)=0$ inward. Since this is a first order equation for $p$, we can see that $R$ is already tied with $M$ so there is no shooting parameter $k$ like in the previous subsection, where $R$ can be set at an arbitrary value as long as $R>2GM$ and $p'(r=0)\sim0$. From $g(r)$, we can see that the equation has singularity at $r=0$ so we expect uncontrolled behavior of $p$ near $r=0$.
	
	\begin{figure} 
		\centering
		\includegraphics[width=0.5\linewidth]{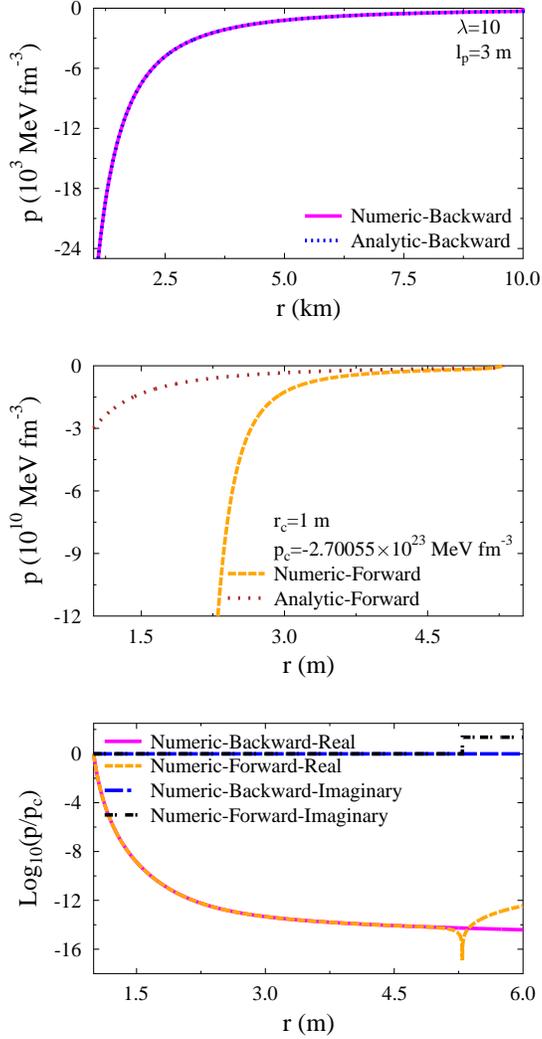}
		\caption{Profiles of $p(r)$ from the positive branch from backward integration (top panel), forward integration (middle panel), and the comparison between each numerical results (bottom panel).Here we calculate first using backward integration, then the values from backward integration is used for forward integration. In the most bottom panel, we show the real and imaginary part of $\log(p/p_c)$ since for the forward integration, $p/p_c=0$ at $r=r_f\sim 5$ m and after this point $p/p_c<0$ when $r>r_f$ and thus $\log(p/p_c)\in\mathbb{C}$. Hence both scheme does not give similar result. 
	     }
		\label{fig:plotprofilCarballoRubio}
	\end{figure}
	
We show the backward integration solutions in top panel of Fig.~\ref{fig:plotprofilCarballoRubio} and it is clear that it violates $|p'(r\to 0)|<\infty$. We tried to use the smallest possible $r=r_c$ as the initial position and its corresponding $p(r_c)$ for starting the forward integration. The resulting curve in the middle panel, while the comparison is in the bottom panel. The results from backward and forward integration are different.  
The real and imaginary words in the most bottom panel is the real and imaginary part of $\log(p/p_c)$ from both backward and forward integration. We show them here since the pressure $p/p_c$ from forward integration suddenly goes to zero at $r=r_f$ much earlier than $R$ from the backward integration, where after this point $p/p_c<0$ and thus $\log(p/p_c)\in\mathbb{C}$ for forward integration. In Fig.~\ref{fig:plotprofilCarballoRubio}, $r_f\sim 5$ m while $R=10$ km.
We still do not yet know how to remedy this, but the numerical result from backward integration perfectly fits the analytic solution \eqref{eq:exactp}.

\subsubsection{\textcolor{black}{Comparison of our numerical results with the density for regular geometry part of Ho-Matsuo model}}
\begin{figure} 
		\centering
		\includegraphics[width=0.9\linewidth]{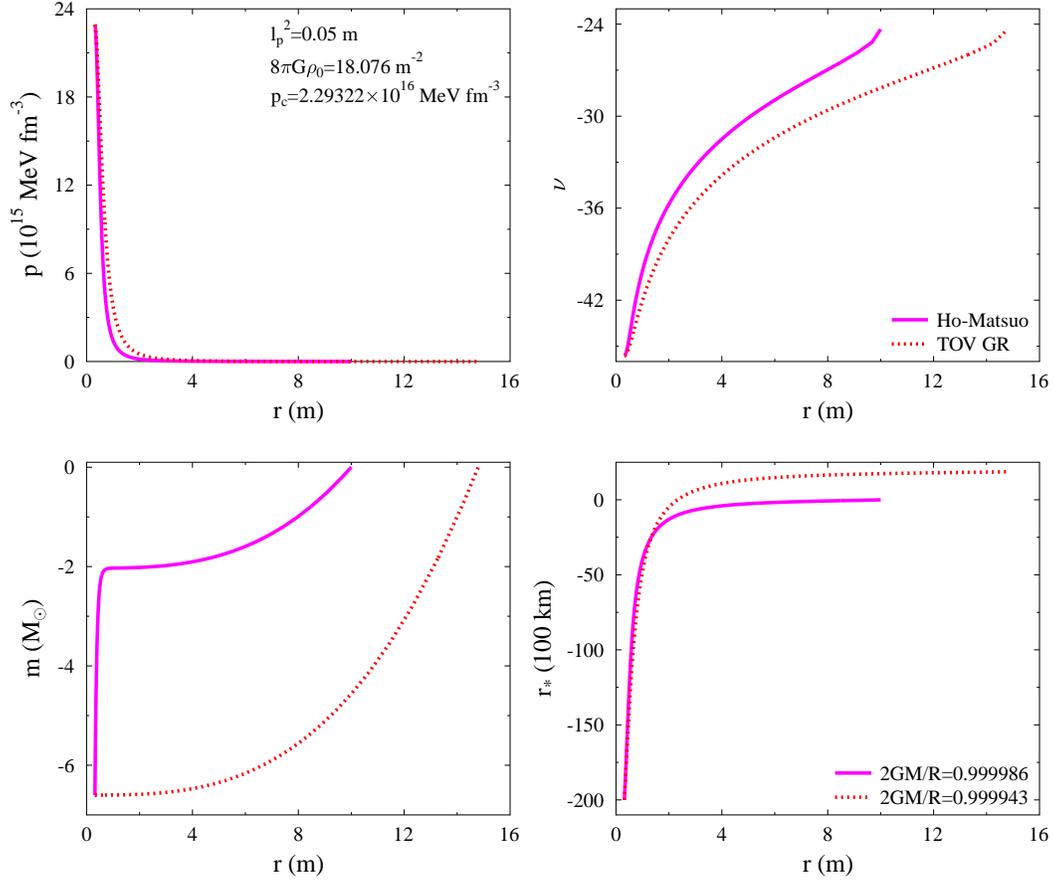}
		\caption{The solid line in these plots are the same as in Fig. \ref{fig:plotprofilhomatsuo} but the dashed line is from TOV GR system. Both data are integrated from center to surface and notice the large compactness $2GM/R$ at the right bottom panel. Notice that $p$ increases dramatically as $r\to 0$. Since in SCGrav $r>l_p$, then $p$ will never go to infinity. But in TOV GR, $r$ can be arbitrarily close to zero so $p$ will go to infinity as $r\to 0$. Notice also that the mass at the center ($m_c$) is negative valued. This may be the reason why the compactness can be very close to black hole limit ($2GM/R=1$) for the case of constant energy density EOS.
		}
		\label{fig:plotprofilHoMatsuo_2}
	\end{figure}

\textcolor{black}{In this subsubsection, we intend to check the validity of our numerical calculation by comparing the typical result of section 6.1.3, density for regular geometry in Ref.~\cite{Ho:2017vgi} with the same EOS (constant energy density EOS) input but calculating by using our code where we restricted the calculation only for the interior of the star.Note that we perform a calculation by using backward and forward integrations to check the consistency of the numerical calculation. It means that the geometry is regular for 0 $\le$ r  $\le$ R, and pressure is finite and positive inside the star. Note that we do not investigate the too small density and too large density cases of the Ho-Matsuo model wherein both latter cases  $r_*$ do not always monotonically increase when  $r$ increased. We had reproduced relatively similar plots with Fig. 14 and Fig. 15 of Ref.~\cite{Ho:2017vgi}, and the trend of the plot satisfies the condition discussed in section 6.1.3 of Ref.~\cite{Ho:2017vgi}. The results are shown in Fig. \ref{fig:plotprofilhomatsuo}. However, we unable to reproduce exactly the quantitative result as shown in Fig. 14 and Fig. 15 of Ref.~\cite{Ho:2017vgi} since the authors of Ref.~\cite{Ho:2017vgi} did not specify the units they are using. Therefore, we only guess the initial conditions until we obtain similar behavior of both $r(r_*)$ and $\nu(r_*)$. We choose $R$ and $\rho_0$ suitably (with $2GM$ fixed) such that the profile $p(r_*)$ decrease monotonically as $r_*$ increases. It turns out that this choice gives us $r$ that is increasing monotonically as $r_*$ increases as the one shows in  Ref.~\cite{Ho:2017vgi}, so in principle, we can interchange $r$ and $r_*$. If we replot all functions in Fig. \ref{fig:plotprofilhomatsuo} as functions of $r$ instead of $r_*$, we obtain Fig. \ref{fig:plotprofilHoMatsuo_2}. In Fig. \ref{fig:plotprofilHoMatsuo_2}, we show the numerical calculation result by integrating the equations \eqref{eq:HoMatsuoEq1}-\eqref{eq:HoMatsuoEq2} from center to surface. We compare it with the usual TOV equation in GR. We can notice that $p$ becomes very steep as $r\to 0$ since $r$ (as a function of $r_*$) drop really quick in that area. We also notice that $m_c$ (mass near the center) is negative valued and quite large in magnitude. We suspect that this negative $m_c$ may be why the compactness can be very close to the black hole limit ($2GM/R=1$). We suspect that this $m_c<0$ is the reason for such high compactness. However, this value is indeed outside the range from our lower and upper bound ($m_{c,min}$ and $m_{c,max}$ from Eq.~\eqref{eq:mcmin} and Eq.~\eqref{eq:mcmax}) which is $-10^{-3} M_\odot\lesssim m_c \lesssim -10^{-7} M_\odot$. Therefore, it seems that our analytical upper and lower bound estimation of $m_c<0$ may not be to justify for constant energy density EOS.
	}
	
\begin{figure} 
		\centering
		\includegraphics[width=0.5\linewidth]{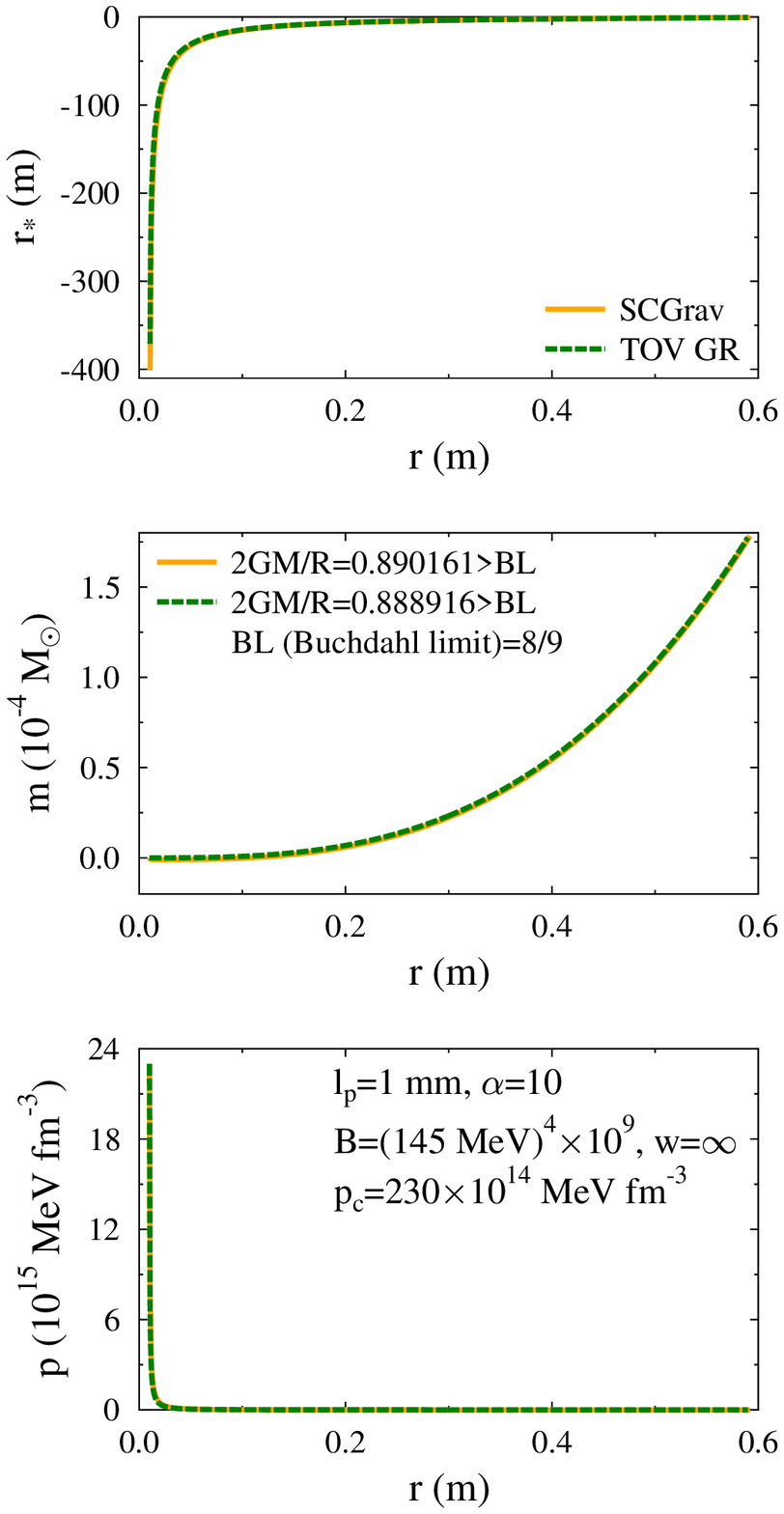}
		\caption{Here we use linear EOS with $w=\infty$, $B \sim 10^{12}$ MeV fm$^{-3}$, and $p_c \sim 10^{16}$ MeV fm$^{-3}$ while $m_c=0$. Unlike Fig. \ref{fig:plotprofilHoMatsuo_2}, the compactness is still far from black hole limit, although is larger than Buchdahl limit (BL).
		}
		\label{fig:verifikasirstar}
	\end{figure}

	\textcolor{black}{To obtain such high compactness, we also observe that the system should have very large density $\rho_0$ and central pressure $p_c$. In our units, we estimate that $\rho_0 \sim 10^{12}$ MeV fm$^{-3}$. This is about nine orders of magnitude larger than the bag constants $B$ that we used. Notice also from Fig. \ref{fig:plotprofilHoMatsuo_2} that $p_c \sim 10^{16}$ MeV fm$^{-3}$. Now we input $\rho_0$ and $p_c$ around this estimation, but with $m_c=0$, into our equations \eqref{eq:m'}-\eqref{eq:eom-end} and integrate them from $r=\alpha l_p$ to $r=R$. The result is shown in Fig. \ref{fig:verifikasirstar}. Note here that $r_*$ is also monotonically increasing with respect to $r$. We also have the compactness slightly above the Buchdahl limit (BL) $2GM/R=8/9$. 
	}

	\textcolor{black}{To verify our expectation that a very high uniform density and central pressure can give us a star with compactness beyond BL, we calculate similar profiles as Fig. \ref{fig:verifikasirstar} with variations of $w$, $B$, and $p_c$. We show the results in Table \ref{table:compactness}. In the top three rows, we vary $w$. In the middle three rows, we vary $p_c$. Moreover, in the top three rows, we vary $B$. We can see that increasing $w$ will increase the compactness. However, to obtain compactness over the Buchdahl limit, we need to increase $B$ to at least eight orders of magnitude and the pressure to at least thirteen orders of magnitude. The entries with bold fonts are from Fig. \ref{fig:verifikasirstar}. We also compare them with the result from TOV GR and find the compactness quite close to SCGrav. Of course, the results from TOV GR that violate the Buchdahl limit will have $p(r\to 0)\to\infty$. Of course, the results from TOV GR that violate the Buchdahl limit will have $p(r\to 0)\to\infty$, which is not correct for the semiclassical model since it should satisfy $r>l_p>0$ condition. We verify this expectation by showing Table \ref{table:compactness}.}
	\begin{table}
		\centering
		\caption{Here is the comparison of compactness with variations of $w$, $B$, and $p_c$, with $m_c=0$ in our semiclassical model (SCGrav) using constant density EOS.}
		\label{table:compactness}
		\begin{tabular}{ccccccc}
			\hline
			$l_p\text{(m)}$ & $\alpha$  & $\left.p_c\text{(MeV }\text{fm}^{-3}\right)$ & $w$ & $\left.B(\text{MeV}^4\right)$ & $2GM/R \text{(SCGrav)}$ & $2GM/R \text{(TOV GR)}$ \\
			\hline
			0.001 & 10 & 230 & \textcolor{blue}{$\text{1/3}$} & $(145)^4$ & $0.531671<\text{BL}$ & $0.528683<\text{BL}$ \\
			0.001 & 10 & 230 & \textcolor{blue}{1} & $(145)^4$ & $0.662569<\text{BL}$ & $0.658682<\text{BL}$ \\
			0.001 & 10 & 230 & \textcolor{blue}{$\infty$}  & $(145)^4$ & $0.754262<\text{BL}$ & $0.74993<\text{BL}$ \\
			0.001 & 10 & \textcolor{blue}{$230\times10^{12}$} & $\infty$  & $(145)^4\times10^9$ & $0.888742<\text{BL}$ & $0.888769<\text{BL}$ \\
			0.001 & 10 & \textcolor{blue}{$230\times10^{13}$} & $\infty$  & $(145)^4\times10^9$ & \textcolor{red}{$0.888961>\text{BL}$} & \textcolor{red}{$0.888902>\text{BL}$} \\
			\textbf{0.001} & \textbf{10} & \textbf{\textcolor{blue}{$230*10^{14}$}} & \textbf{$\infty$}  & \textbf{$(145)^4*10^9$} & \textbf{\textcolor{red}{$0.890161>\text{BL}$}} & \textbf{\textcolor{red}{$0.888916>\text{BL}$}} \\
			0.001 & 10 & $230*10^{14}$ & $\infty$  & \textcolor{blue}{$(145)^4*10^8$} & \textcolor{red}{$0.889297>\text{BL}$} & \textcolor{red}{$0.888892>\text{BL}$} \\
			0.001 & 10 & $230*10^{14}$ & $\infty$  & \textcolor{blue}{$(145)^4*10^7$} & \textcolor{red}{$0.889019>\text{BL}$} & $0.888889=\text{BL}$ \\
			0.001 & 10 & $230*10^{14}$ & $\infty$  & \textcolor{blue}{$(145)^4*10^6$} & \textcolor{red}{$0.88893>\text{BL}$} & $0.888889=\text{BL}$ \\
			\hline
		\end{tabular}
	\end{table}


\end{document}